# An Immersive Visualization Tool for Teaching and Simulation of Smart Grid Technologies

[1]Dr. J. Chris Foreman, [2]Dr. Rammohan K. Ragade, and [1]Dr. James H. Graham
University of Louisville Speed Scientific School of Engineering
[1]Deptartment of Electrical and Computer Engineering
[2]Deptartment of Computer Engineering and Computer Science
Louisville, KY 40292

**Abstract** – Intelligent power grid research, i.e. smart grid, involves many simultaneous users spread over a relatively large geographical area. A tool for advancing research and community education is presented utilizing large-scale visualization centers, e.g. planetariums, in simulating smart grid interactions. This approach immerses the user in virtual smart grid visualization and allows the user, with several other users, to interact in real time. This facilitates community education by demonstrating how the power grid functions with smart technologies. The simulation is sophisticated enough to also be used as a research tool for industry and higher education to test software algorithms, deployment strategies, communications protocols, and even new hardware.

## I. INTRODUCTION

The *smart grid* has become a popular topic in research and media. There are several aspects that have been discussed such as demand response, smart metering, technologies for enabling renewable energy, and so on. The general public, most of whom are not familiar with our existing power grid infrastructure, are often left wondering what this really means to them. Researchers are often focused on a particular area and though there is much collaboration, there is need for a system-level approach that comprehensively includes the existing power grid, these new smart grid technologies, and the diverse users who will utilize, research, and learn from them together, in real time.

The proposed immersive visualization tool (IVT) seeks to provide a simulation capable of supporting both existing and proposed power grid technologies while immersing users within a representative visualization environment. This IVT allows several users to interact with the simulation as individuals while observing the effects to the power grid system as a whole. Observing in this comprehensive manner allows new aspects of these technologies and their cross impacts to be discovered. From an educational standpoint, the users are educated in the functionality of the power grid and new smart grid technologies. From the research and development standpoints, users are researchers, power companies, etc testing new smart grid technologies such as strategies, algorithms, grid reconfiguration, and power electronics hardware, within a near real-world environment. The IVT becomes a foundation in which these new technologies are developed, and one in which such technologies can be demonstrated to the public before deployment. A model of the components of the IVT and the types of users are illustrated in Fig. 1. These will be discussed further in section II.

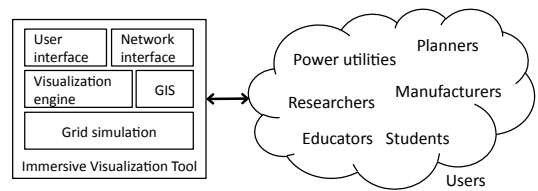

Fig. 1 – The IVT and its users.

The design of the IVT is capable of hierarchical application layers. Fig. 2 illustrates how increasing abstraction provides a path from low-level equipment and automation up through high-level management strategies. In this way, the IVT truly becomes a tool for smart grid utilization.

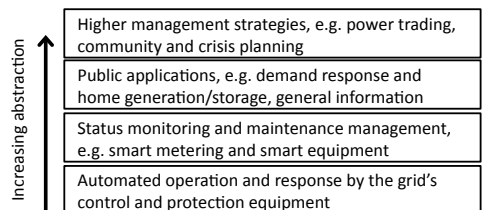

Fig. 2 – Application layers.

*A. Current research*

There is need for better power grid visualizations, particularly for the education of new power grid professionals, which in itself is a whole topic [1] [9]. Better visualization results in better situation awareness and this will only become more necessary as power grids evolve into smart grids [2]. There are also concerns with public acceptance of these new technologies that they may interfere with electricity delivery or increase the costs to consumers [12].

Visualization systems that are coupled with geographical information systems (GIS) enable many types of power grid data to be viewed, such as power flows, capacity, trouble conditions, weather, etc by a wider range of users such as power traders, grid managers, planners, and the general public [3]. There has been work in incorporating power grid support into traditional GIS, particularly in handling the spatio-temporal data associated with power simulations versus the spatial-only data typical of geographic data [4]. Other efforts for advanced visualization techniques include those for high-speed data visualization, e.g. frequency [5], and even techniques resulting in a collaborative 3-dimensional information platform for various users [6].

*GreenGrid* is another detailed visualization system for power grid analysis developed collaboratively for researchers by the Department of Energy (DOE) and the Pacific Northwest National Laboratory (PNNL) centers for National Visualization and Analytics (NVAC) and Electricity Infrastructure Operations (EIOC). [7] Visualizing Energy Resources Dynamically on Earth (VERDE) is another innovative effort using Google Earth by Oak Ridge National Laboratory [13]. There have even been efforts in multiuser game environments that seek to educate the general public such as World of Warcraft and SimCity [11].

In the above efforts, work is done to enable power grid analysis in a GIS-type environment and to make the work available for researchers. Power utilities also use visual approaches, as in Fig. 3, but these are limited to system operation and maintenance. In neither of these cases, however, is an immersive visualization developed that is comprehensive, as in Fig. 1, for all these users and smart grid technologies.

To realize the immersive aspect in this IVT, planetariums are utilized due to their ability to seat larger audiences and provide an encompassing display. Planetarium visualizations put the user (observer) into the visualization and are rapidly becoming a pinnacle of scientific visualization [8]. The ability of the users to interact in real time with the IVT facilitates learning from the elementary level, up through advanced concepts for students of higher education. This also provides a framework for testing new strategies on the power grid, e.g. demand response, by researchers and industry professionals to evaluate cross impacts, not just analytically but socially as well.

## II. APPROACH

The goals of the IVT are to realize a tool capable of an engaging simulation of the smart grid for community education, research, and planning. To achieve these goals, the IVT is designed with the following qualities:
- Engaging, immersive – this keeps the focus of the audience and researchers and allows them to relate to the IVT in a meaningful way.
- Expandable, adaptable – this allows the IVT to adapt to changes in smart grid topics and expand as technologies are added.
- High resolution, accurate – the IVT is not only for generalized demonstration but also needs to be granular to allow fine details to be explored accurately enough for real world implementation.

A detailed repository is being developed by coordinated efforts at the National Institute of Standards and Testing (NIST) of smart grid use cases [14]. The users of the IVT have multiple points of view and therefore the tool is utilized to assemble the following use case groupings:
- Education – This group includes the general community as well as secondary and higher education. Existing power grid principles and smart grid technologies can be demonstrated interactively to this group via the IVT. Users in this group can also *play* the IVT as a game scenario to learn interactions.
- Researchers and developers – This group includes theoretical research, product development, and testing. The IVT is used to simulate various technologies in a near real world environment to see how they work together, i.e. cross impact studies.
- Utility companies – This group includes power utility companies. In addition to developing operational and maintenance strategies, corporate strategies such as power trading, new service deployments, etc can be

- explored, and even demonstrated to the public, before deployment using the IVT.
- Community planners – This group includes public service commissions, local governments, and response teams. They also seek to evaluate new services and explore crisis response, but from a regulatory point of view. Other impacts such as environmental, economic, and social can be explored with the IVT.

GIS systems utilize a layered approach. The IVT is also developed in layers starting with a community map. Additional object layers of power consumers, power producers, power grid equipment, etc can then be incrementally laid on top. Likewise, soft technologies such as demand response, smart metering, power trading, etc is layered in as well via scripting to intelligently link these objects.

*A. Building the power grid display*

The first step is to start with the power grid configured as it is today. Power grid visualization is already done by power utilities to monitor the local grid and manage faults for repair crews. Some of these displays are large, but they are not designed for ease of use beyond the power utilities' distribution personnel. They are also not designed for research and collaborative access. Fig. 3 illustrates a typical power distribution display of a local utility.

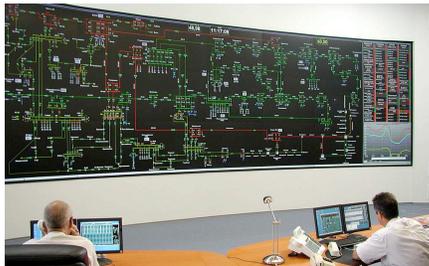

Courtesy SO-CDO UES Russian Power Grid
Fig. 3 – Typical utility power grid display.

Utilizing a GIS system, a satellite image of the local community can be displayed. The power grid components are then added as a layer in cooperation with the local power utility. This allows users to observe the geographical layout of the power grid. When community users such as homeowners and businesses observe the IVT, they can see their location within the power grid.

Operational data and interaction are the next key elements to integrate into the IVT. Operational data can be simulated from algorithms that define devices scripted into the GIS software or loaded as live data from field I/O via an Internet gateway. Interaction is integrated by users interfacing with software agents that represent entities/devices on the power grid. For example, a software agent simulates a single residential user's home. The user can then interact with this agent to observe the effects throughout the power grid of various actions. These actions include appliance settings, configuration of onsite power generation, onsite power storage, demand response, etc. Many of these agents acting in parallel results in an interactive model of the power grid, and one in which multiple users can utilize simultaneously. When the IVT is being displayed, multiple users can access specific devices and make changes to that device's power profile. Fig. 4 demonstrates two users interacting with their homes and observing the cross impacts of their actions.

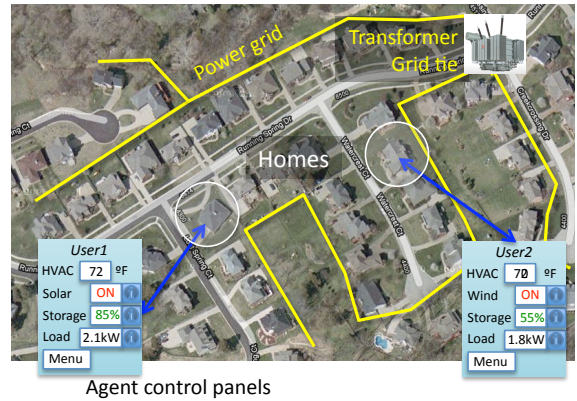

Fig. 4 – Software agents representing homes.

*B. Using planetariums for visualization*

While a simulation is useful on a single computer screen, a large-scale display is required to observe at a more granular level, interact with multiple users, and present to larger audiences. Metropolitan areas and power grid jurisdictions are geographically spread out, yet individual devices, such as a single transmission line or power transformer, need to be observed in detail within this large-scale view. This is only possible with a visualization-scale display. Appropriate seating is also necessary to handle several users from collaborative research teams to public audiences. A planetarium is a natural venue for such visualization [8]. Planetariums supply optimal seating for audiences while affording optimal views of the visualization. The size and resolution of the planetarium display allows large, detailed views.

As a collaborative effort to this work, the University of Louisville Rauch Planetarium is in application for a visualization projector capable of such a display. The Rauch planetarium houses a 55ft dome with a seating capacity of 160 people. Once equipped, full dome visualizations 360º x 180º will be viewable to the audience as illustrated in Fig. 5. In addition to this, Wi-Fi access to the IVT server is provided to allow multiple users to access the IVT. This access can be via a web interface accessed by the user's PDA/PC, e.g. an iPhone. The planetarium is also on the university's high-speed computer network, which links the IVT to on-campus supercomputing capabilities and the Internet. Fig. 6 illustrates the information network available to the planetarium.

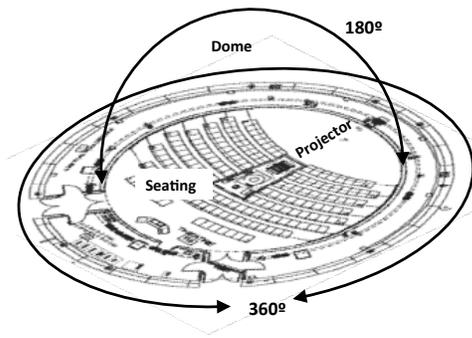

Fig. 5 – Planetarium visualization display (dome).

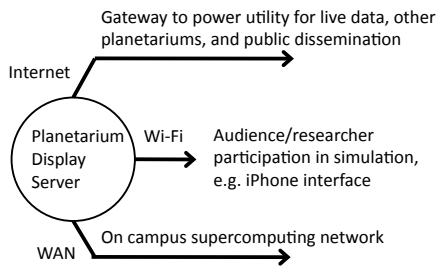

Fig. 6 – Planetarium information network.

*C. Adding smart grid technologies*

Once the foundation is set for simulation and visualization of the power grid via the IVT within a planetarium environment, new technologies for smart grid are then layered in for strategic testing and evaluation. Smart grid technologies include, but are not limited to, the following:

- Demand response – setting preferences for individual locations and observing the system response.
- Distributed generation – as consumers add onsite generation, typically through renewables, the effects of penetration can be observed.
- Distributed storage – consumers and the power utility itself may add power storage at various points and the stabilizing effects can be observed and quantified.
- Other strategies – new technologies and strategies are being proposed all the time. These could be scripted into the IVT and both evaluated and publicly disseminated quickly.

GIS are particularly well suited for smart grid simulation. They usually incorporate a scripting language that facilitates the definition of objects and interactions. They link well with external databases and can therefore incorporate user interactions and operational power grid data as previously mentioned. They also scale well to specialty displays, such as planetarium domes or multi-screen walls. Primarily, they allow objects to be added and related by geographical location. These objects are then linked together in meaningful ways. Fig. 7 illustrates how a transformer, from Fig. 4, is defined using the transformer's characteristics, operational (live) data, and action scripts that define how it behaves and links to other objects.

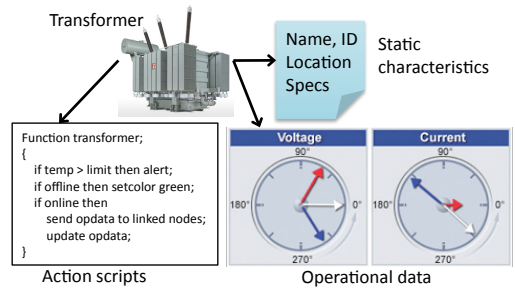

Fig. 7 – GIS definition of transformer object.

For example, roads are linked on maps to determine routes using characteristics such as travel direction and operational data such as current traffic. Similarly, power grid components, distributed generation/storage, and consumers can be defined in this GIS framework to simulate the power grid and with added layers, the smart grid.

*D. IVT project implementation path*

The IVT is developed along a path similar to that illustrated in Fig. 8. It is anticipated that incremental results can be obtained and disseminated along the demonstrated path. Collaborators include University of Louisville Speed Engineering School for the overall IVT development with strategic input from the local power utility EON.US. The Rauch Planetarium will be a key visualization partner under the university's Department of Education, which will also provide guidance in community and academic educational efforts.

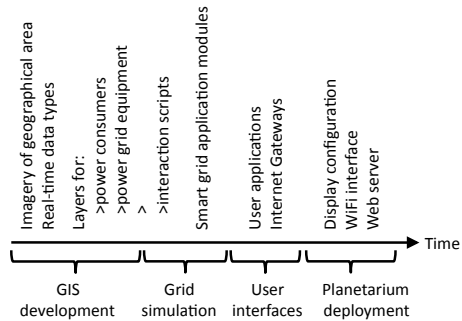

Fig. 8 – Project implementation path.

The IVT is in the proposal and planning stages as of April 2010. It is expected that the *GIS development* and *Grid simulation* stages will be underway during this year and that some of these efforts will be disseminated at the *IEEE SmartGridComm* conference at NIST in October 2010.

III. CONCLUSIONS

The IVT introduced in this paper serves a unique role in which simulation and education aspects can be jointly explored. The tool is designed to be accessible across diverse users of various levels of expertise. The tool is also designed to include many technological layers together and allow cross impact analysis. Finally, the unique display environment and capability of interaction results in a solution that is truly immersive.

*A. Additional opportunities*

Once deployed, the IVT is designed to continue in growth to reach more users and encompass more technological applications. Some aspects that intend to be explored further once the IVT of section II is realized include the following:

- In addition to the power grid, the same principles can be applied to the water, gas, and communications infrastructure. This additional scope can be layered in with other technologies/objects as previously discussed. Fig. 9 illustrates this principle.
- Communications protocol testing could be facilitated so that new standards could be compared in a comprehensive, near real world environment. Error and performance testing could be included.
- Infrastructure security is an area to be explored. The smart grid is likely to close many existing vulnerabilities, but the complexity of new technologies interacting is also likely to introduce new vulnerabilities. Again, the comprehensive environment of the IVT allows these to be quantified and addressed [10].
- Consumer acceptance is a key element to successfully implementing smart grid technologies [12]. It has been mentioned that the IVT, through education, could benefit this. Additionally, a publicly available WWW version, perhaps implemented as a game, could further the impact of this aspect [11].

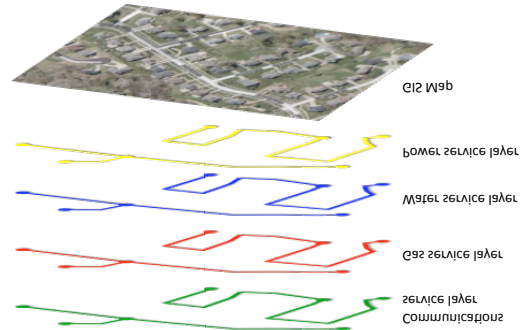

Fig. 9 – Infrastructure layers.

REFERENCES

[1] S. Brahma et al, "The Education and Training of Future Protection Engineers: Challenges, Opportunities, and Solutions," *IEEE Transactions on Power Delivery*, vol. 24, no. 2, pp. 538-544, April 2009.
[2] Overbye, T.J., "Transmission system visualization for the smart grid," *Power Systems Conference and Exposition, 2009*. PSCE '09. IEEE/PES, pp. 1-2, 15-18 March 2009.
[3] Overbye, T.J., Weber, J.D., "Visualizing the electric grid," *Spectrum, IEEE*, vol. 38, no. 2, pp. 52-58, Feb 2001.
[4] Parikh, P.A., Nielsen, T.D., "Transforming traditional geographic information system to support smart distribution systems," *Power Systems Conference and Exposition, 2009*. PSCE '09. IEEE/PES, pp. 1-4, 15-18 March 2009.
[5] Bank, J.N., Omitaomu, O.A., Fernandez, S.J., Yilu Liu, "Visualization and Classification of Power System Frequency Data Streams," *Data Mining Workshops, 2009*. ICDMW '09. IEEE International Conference on, pp. 650-655, 6-6 Dec. 2009.
[6] San Yecai, Zhu Chuanbai, Cao Yijia, Guo Chuangxin, "An architecture of spatial three dimension visualization information platform for urban power grid," *Universities Power*

*Engineering Conference, 2008*. UPEC 2008. 43rd International, pp. 1-5, 1-4 Sept. 2008.

[7] Pak Chung Wong et al, "A Novel Visualization Technique for Electric Power Grid Analytics," *Visualization and Computer Graphics, IEEE Transactions on*, vol. 15, no. 3, pp. 410-423, May-June 2009.

[8] Ryan Wyatt, "Science Visualisation within a Planetarium Setting," *Communicating Astronomy with the Public 2007: Proceedings from the IAU/National Observatory of Athens/ESA/ESO Conference*, Athens, Greece, 8-11 October 2007.

[9] Gerald T. Heydt et al, "Professional Resources to Implement the Smart Grid," *41st North American Power Symposium*, October 4–6, 2009, Starkville, MS USA.

[10] Khurana, H., Hadley, M., Ning Lu, Frincke, D.A., "Smart-Grid Security Issues," *Security & Privacy, IEEE,* vol. 8, no. 1, pp. 81-85, Jan-Feb. 2010.

[11] Jaymi Heimbuch, "World of Warcraft an Unlikely Tool for Environmentalism," World Wide Web found at http://www.treehugger.com/files/2009/02/world-of-warcraft-an-unlikely-tool-for-environmentalists-video.php, Feb 2010.

[12] Richard Feinberg, "Achieving Customer Acceptance of the Smart Grid," *Report by The Intelligent Project LLC*, Purdue Research Park, April 15, 2009.

[13] Thomas King et al, "Visualizing Energy Resources Dynamically on Earth: VERDE: A Real-time National Visualization Resource," Fact Sheet Oak Ridge National Laboratory, Aug 2008.

[14] NIST Smart Grid Collaboration Site, World Wide Web found at http://collaborate.nist.gov/twiki-sggrid/bin/view/SmartGrid/IKBUseCases, Apr 2010.